\documentclass[prl,twocolumn,superscriptaddress,showpacs]{revtex4}
\usepackage{graphicx}
\usepackage{epsfig}
\usepackage{amssymb,amsmath,amsfonts,hyperref}
\usepackage{wasysym}
\usepackage{latexsym}
\usepackage{eucal}
\usepackage{verbatim}
\usepackage[normalem]{ulem}
\usepackage{times}

\usepackage{color}

\newcommand{\be}{\begin{eqnarray}}
\newcommand{\ee}{\end{eqnarray}}

\begin{document}
\title{Phase transitions in the distribution of the Andreev conductance of superconductor-metal
junctions with multiple transverse modes.}
\author{Kedar Damle}
\affiliation{Tata Institute of Fundamental Research, 1, Homi Bhabha Road,
Mumbai 400005, India}
\author{Satya N. Majumdar}
\affiliation{Univ. Paris-Sud, CNRS, LPTMS, UMR8626, Orsay F-01405, France}
\author{Vikram Tripathi}
\affiliation{Tata Institute of Fundamental Research, 1, Homi Bhabha Road,
Mumbai 400005, India}
\author{Pierpaolo Vivo}
\affiliation{Univ. Paris-Sud, CNRS, LPTMS, UMR8626, Orsay F-01405, France}

\begin{abstract} 
We compute analytically the full distribution of Andreev conductance $G_{\mathrm{NS}}$ of
a metal-superconductor interface with a large number $N_c$ of transverse modes, using a random matrix approach.
The probability distribution $\mathcal{P}(G_{\mathrm{NS}},N_c)$ in the 
limit of large $N_c$ displays a Gaussian behavior
near the average value $\langle G_{\mathrm{NS}}\rangle= (2-\sqrt{2}) N_c$ and 
asymmetric power-law tails in the two limits of 
very small and very large $G_{\mathrm{NS}}$. 
In addition, we find a novel third regime sandwiched between the central
Gaussian peak and the power law tail for large $G_{\mathrm{NS}}$.
Weakly non-analytic points separate these four regimes---these are shown
to be consequences of
three phase transitions in an associated Coulomb gas problem.
\end{abstract}

\vskip2pc

\maketitle

\textit{Introduction} - Advances in fabrication of mesoscopic structures has
led to a great deal of interest in their electrical and thermal
transport properties, from the point of view of both fundamental
questions in the quantum theory of transport, and of device applications \cite{Mesoscopics}. When the devices are disordered or chaotic, a statistical 
approach in which one characterises the phase-coherent
motion of electrons in terms of 
an ensemble of unitary scattering matrices $\mathbf{S}$ \cite{Imry,Muttalib_Pichard_Stone,Mello,Baranger_Mello,Jalabert_Pichard_Beenakker,RMTreview} and uses
Landauer's description \cite{landauer,Fisher_Lee} of transport in terms of the corresponding transmission eigenvalues $\{T_n\}$, has proved very successful. 
Among the early successes of this approach was a general
and transparent explanation \cite{Imry,Muttalib_Pichard_Stone,Mello} for the phenomenon of \emph{universal conductance fluctuations} \cite{Alt,Lee_Stone,Mesoscopics}:
the variance $\text{var}(G)$ corresponding to
sample-to-sample fluctuations of the 
conductance $G$ (measured in
units of the conductance quantum $G_0 = 2e^2/h$) of disordered mesoscopic
structures is independent of their size and the disorder
strength, and is determined
solely by whether or not time-reversal (TR) and other symmetries are present.

Within this random matrix approach, the conductance $G$ of a
structure with $N_c$ transverse channels is given
as $G = \sum_{n=1}^{N_c} T_n$,
and the fact that its variance $\text{var}(G)$ is a universal ${\mathcal O}(1)$ number
is then seen to be a natural consequence of {\em strong
correlations} between the $\{T_n\}$---the precise nature of these
correlations is determined only by the symmetry properties of the relevant ensemble of scattering matrices.
These correlations cause
$\text{var}(G)$ to become independent of
$N_c$ at large $N_c$, contrary to expectations from the usual
`central limit considerations' for sums of a large number of {\em independent} random variables.

How do these strong correlations affect the form of the {\em full probability
distributions} of various transport properties, including their large
deviations from the mean? This question is interesting not only
because recent experimental advances may make it possible to 
{\em measure} such distribution functions in
some cases \cite{hemmady,Fridman},
but also because similar questions about the behaviour of correlated random variables have recently surfaced 
in many disparate fields with a large number of applications \cite{Touchette}.
In spite of this broad interest, there are few results available
along these lines---notable among these are
the recent calculations for the full distribution of the conductance
and shot-noise of mesoscopic structures in their \textit{normal} metallic state \cite{kanz,kanz2,Vivo,Vivo_Majumdar_Bohigas}, and for chaotic structures with one or two superconducting outgoing channels \cite{gopar}.

In this Letter, we have obtained the full distribution of the conducance $G_{\mathrm{NS}}$ of a time-reversal symmetric normal metal-superconductor (NS) junction in the limit of large $N_c$. Transport across an NS junction is particularly
interesting because an electron incident from the normal side
can be reflected as a hole, with the injection of a Cooper pair into
the superconducting condensate \cite{andreev}. Incorporating
the effects of such processes in the presence of TR
symmetry allows one to write the conductance $G_{\mathrm{NS}}$ (measured
in units of $G_0$) of such
junctions as $ G_{\mathrm{NS}} = 2\sum_{n=1}^{N_c}\left(\frac{T_n}{2-T_n}\right)^2$, where $\{T_n\}$ are the transmission eigenvalues of the same
junction in its putative normal state \cite{been}.
The conductance $G_{NS}$ 
thus ranges from $0$ to $2N_c$, and its average $\langle G_{\mathrm{NS}} \rangle=(2-\sqrt{2})N_c$ and variance $\mathrm{var}(G_{\mathrm{NS}})=9/16\simeq 0.563$
are well-known in this TR symmetric case \cite{Jalabert_Pichard_Beenakker} (see also \cite{footer}).

\begin{figure*}
 {\includegraphics[width=8.7cm]{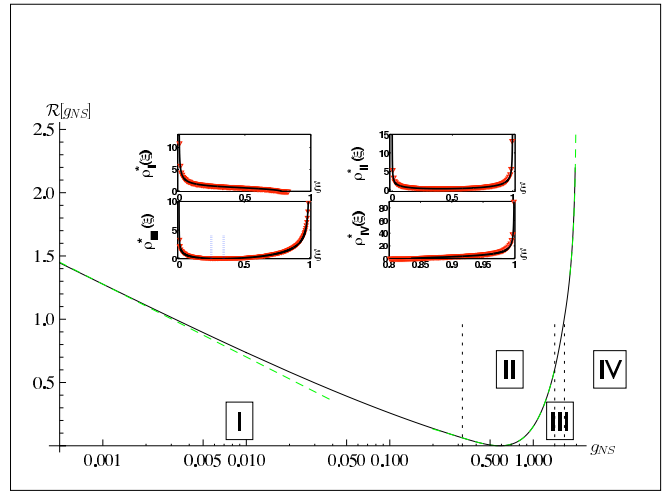}}
\caption{The rate function $\mathcal{R}(g_{\mathrm{NS}})$ obtained from our large $N_c$
solution is shown along with an inset that displays the form of the equilibrium density
of the Coulomb gas in each regime (analytical formulae in solid black lines and Monte Carlo simulations in red triangles, see \cite{supp} for details). The green dashed lines are fit
to the asymptotic forms for the left and the right tails 
and the central Gaussian region mentioned
in the text. The vertical black dashed lines correspond to the critical points $g_1,g_2$ and $g_3$.} 
\label{Fig1}
\end{figure*}

Here we show that $\mathcal{P}(G_{\mathrm{NS}},N_c)$ for large 
$N_c$ has the scaling form \cite{footnote2}:
\begin{equation}
\mathcal{P}(G_{\mathrm{NS}},N_c) 
\approx \exp\left(-N_c^2 \mathcal{R}(g_{\mathrm{NS}})\right),
\label{largedeviationansatz}
\end{equation}
where the large deviation function $\mathcal{R}(g_{\mathrm{NS}})$ is plotted in Fig. \ref{Fig1}
and $g_{\mathrm{NS}}\in [0,2]$ is the dimensionless conductance per channel, $g_{\mathrm{NS}}= G_{\mathrm{NS}}/N_c$ . A striking
consequence of our exact computation of $\mathcal{R}$ is the
prediction of a marked asymmetry in the large-deviation asymptotics
near $G_{\mathrm{NS}}\rightarrow 0$ where $\mathcal{P}(G_{\mathrm{NS}}, N_c) \sim  g_{\mathrm{NS}}^{N_c^{2}/4}$ and near $G_{\mathrm{NS}} \rightarrow 2N_c$
where $\mathcal{P}(G_{\mathrm{NS}},N_c) \sim  (2-g_{\mathrm{NS}})^{N_c^{2}/2}$. 

Another interesting feature is that the rate function $\mathcal{R}(g_{\mathrm{NS}})$ 
is piecewise smooth over a domain consisting of {\em four} distinct regions $g_{\mathrm{NS}}\in\bigcup_{j=0}^3 [g_j,g_{j+1}]$
glued together via weak non-analytic points: apart
from the {\em asymmetric} large-deviation tails displayed above for
$g_{\mathrm{NS}}$ near $0$ and $2$ respectively, and the universal Gaussian
behavior
\begin{equation}
\mathcal{P}(G_{\mathrm{NS}},N_c) \sim \exp\left(-(G_{\mathrm{NS}} -\langle G_{\mathrm{NS}}\rangle)^2/2\sigma^2\right),
\label{gaussian}
\end{equation}
with dimensionless variance $\sigma^2 = 9/16$ around the mean $\langle G_{\mathrm{NS}}\rangle = (2- \sqrt{2})N_c$, there is a 
fourth tiny region that separates
the Gaussian central region from the large-deviation tail near $G_{\mathrm{NS}} = 2N_c$. As we shall demonstrate below, this
is a direct consequence of three phase transitions in an associated Coulomb gas problem.

\textit{The Coulomb gas problem} -
The $N_c$ transmission eigenvalues $T_n\in [0,1]$ are distributed according to
the Jacobi Orthogonal random matrix ensemble \cite{RMTreview}:
\begin{equation}
\mathcal{P}_{\mathbf{T}}\left(\{T_n\}\right) = 
A_{N_c} \prod_{n < m}|T_n - T_m| \prod_{n}T_n^{-1/2},
\label{jointpdfofT}
\end{equation}
with $A_{N_c}$ ensuring normalization.
The probability distribution of $G_{\mathrm{NS}}$ is given by $\mathcal{P}(G_{\mathrm{NS}}, N_c)=$
\begin{equation}
=\int_{[0,1]^{N_c}}\prod_i dT_i\delta\left(g_{\mathrm{\mathrm{\mathrm{\mathrm{NS}}}}}N_c-2\sum_{n=1}^{N_c}\frac{T_n^2}{(2-T_n)^2}\right)\mathcal{P}_{\mathbf{T}}\left(\{T_n\}\right). \label{formaldefinition}
\end{equation}
Changing variables $\xi_n = T_n/(2-T_n)$ and exponentiating the $\delta$ function \cite{supp} leads to $\mathcal{P}(G_{\mathrm{NS}}, N_c)=$
\begin{equation}
=\frac{N_c}{2}\int\frac{d\kappa}{2\pi}\int_{[0,1]^{N_c}}\prod_i d\xi_i\ e^{\mathrm{i}N_c^2\kappa\left(\frac{1}{N_c}\sum_{n=1}^{N_c}\xi_n^2-\frac{g_{\mathrm{NS}}}{2}\right)}\mathcal{P}_{\mathbf{\xi}}\left(\{\xi_n\}\right),
\label{partitionsum}
\end{equation}
where
\begin{equation}
\mathcal{P}_{\mathbf{\xi}}\left(\{\xi_n\}\right)=
\tilde{A}_{N_c}\prod_{n < m}|\xi_n - \xi_m| \prod_{n}\frac{\xi_n^{-1/2}}{(1+\xi_n)^{N_c+1/2}}
\label{jointpdfofxi}
\end{equation}
and $\kappa$ is constrained by the saddle-point condition to be purely imaginary \cite{supp}.
While the $N_c$-fold $\{\xi\}$ integral \eqref{partitionsum} can be computed for any finite $N_c$ in terms of Pfaffians \cite{details}, 
for large enough $N_c$ one can map \eqref{partitionsum} to a continuum Coulomb gas problem. 
To make this connection, we represent a particular
realization of $\xi$ in terms of a continuum density function $\rho(\xi) =\frac{1}{N_c}\sum_{n=1}^{N_c}\delta(\xi - \xi_n)$
obeying the normalization condition $\int_0^1d\xi \rho(\xi) = 1$. Originally introduced by Dyson~\cite{Dyson}, this procedure
has recently been successfully used in a number of different contexts \cite{DM,Vivo_Majumdar_Bohigas_JMathPhys,Majumdar_Nadal_Scardicchio_Vivo}.

We may now write the probability distribution $\mathcal{P}(G_{\mathrm{NS}},N_c)$
in this large $N_c$ limit
as a functional integral over the normalized density field $\rho$,
supplemented by two additional integrals enforcing two constraints
\begin{equation}
\mathcal{P}(G_{\mathrm{NS}}, N_c) = {\cal A}_{N_c}\int dC_0 \int dC_1 \int {\cal 
D}\rho \exp\left(-N_c^2\mathcal{S}[\rho]\right),
\label{functionalintegral}
\end{equation}
where the action $\mathcal{S}$ is given by
\begin{widetext}
\begin{equation}
\mathcal{S}[\rho] = C_1 \left( \int d\xi \xi^2 \rho(\xi)- \frac{g_{\mathrm{NS}}}{2} \right)  + C_0\left(\int d\xi \rho(\xi) - 1\right) +\int d\xi \rho(\xi)\ln(1+\xi) - \frac{1}{2} \int \int d\xi d\xi^{'} \rho(\xi)\rho(\xi^{'})\ln|\xi - \xi^{'}|. \label{action}
\end{equation}
\end{widetext}
Here,  ${\cal A}_{N_c}\sim\exp(N_c^2\Omega_0)$, 
with $\Omega_0=(3/2)\ln 2$ is the overall normalization factor
in this large $N_c$ limit. The two variables $C_0$ and $C_1=-\mathrm{i}\kappa$
represent the integral representations of the two delta functions enforcing
respectively the 
normalization condition  $\int_0^1d\xi \rho(\xi) = 1$ 
and $ \int d\xi \xi^2 \rho(\xi)=\frac{g_{\mathrm{NS}}}{2} $. We have also dropped 
contributions to the action $\mathcal{S}$ that
are subdominant in the large $N_c$ limit. For notational convenience,
we have also suppressed the $C_0$ and $C_1$ dependence of the action $S[\rho]$. 

Clearly, \eqref{functionalintegral} can be viewed as
the partition function of a 2-d gas of particles confined on the 
segment $[0,1]$, subject to an all-to-all Coulomb repulsion
and sitting in an external potential
$V(\xi)=\ln(1+\xi) + C_1\xi^2 + C_0$ at inverse temperature $N_c^2$.
In this large $N_c$ limit, equilibrium properties of this Coulomb
gas are clearly determined by the saddle point of the functional 
integral \eqref{functionalintegral},
that corresponds to the minimum energy configuration of the fluid.
We have three saddle points equations. Varying $\mathcal{S}[\rho]$ over $C_0$ and 
$C_1$
just give the two constraints mentioned above. The third equation 
$\frac{\delta \mathcal{S}[\rho]}{\delta \rho}=0$, gives 
the 
minimum energy
density configuration $\rho^{\star}$ which satisfies the integral equation
\begin{equation}
\ln(1+\xi) +C_0 + C_1\xi^2 = \int \rho^{\star}(\xi^{'})\ln|\xi - \xi^{'}|d\xi^{'}
\label{integralequation}
\end{equation}
for all $\xi$ in the support of $\rho^{\star}$. Differentiating \eqref{integralequation} with respect to $\xi$ we get
\begin{equation}
2C_1\xi + \frac{1}{1+\xi} = \Pr \int \frac{\rho^{\star}(\xi^{'})}{\xi - \xi^{'}}d\xi^{'}
\label{differentiatedintegralequation}
\end{equation}
for all $\xi$ in the support of $\rho^{\star}$, where $\Pr$ stands for Cauchy's principal part. 

Finding the solution $\rho^{\star}(\xi)$ of \eqref{differentiatedintegralequation}
with the constraints $\int_0^1d\xi \rho^\star(\xi) = 1$ and
$\int_0^1d\xi \xi^2\rho^\star(\xi) = g_{\mathrm{NS}}/2$ is the main technical challenge.
The saddle
point density $\rho^{\star}(\xi)$ obtained in this manner then
depends parametrically only on $g_{\mathrm{NS}}\in [0,2]$, and the required result for
the probability distribution in the large $N_c$ limit is finally given in terms of the action $\mathcal{S}$
evaluated on $\rho^{\star}$,
\begin{equation}
\mathcal{P}(G_{\mathrm{NS}}, N_c) \approx \exp\left[-N_c^2 \underbrace{\left(\mathcal{S}[\rho^{\star}]-\Omega_0\right)}_{\mathcal{R}(g_{\mathrm{NS}})}\right].
\label{result}
\end{equation}

\textit{Solution of \eqref{differentiatedintegralequation} and phase transitions for $\rho^{\star}$} - 
Singular integral equations of the type \eqref{differentiatedintegralequation} can be solved in closed form using either Tricomi's theorem \cite{Tricomi} when $\rho^{\star}$ has support
on a {\em single interval} $[L_1,L_2]$, or a more general scalar Riemann-Hilbert method \cite{Brezin,Vivo_Majumdar_Bohigas} if this assumption
is not valid. 

We find that \cite{details}
\begin{equation*}
\rho^{\star}(\xi)=
\begin{cases}
\rho^{\star}_{I} (\xi) &\mbox{for } g_0=0 \leq g_{\mathrm{NS}} \leq g_1, \quad\mbox{see }\eqref{solution1}\\
\rho^{\star}_{II} (\xi) &\mbox{for } g_1 \leq g_{\mathrm{NS}} \leq g_2, \quad\mbox{see }\eqref{solution2}\\
\rho^{\star}_{III} (\xi) &\mbox{for } g_2 \leq g_{\mathrm{NS}} \leq g_3, \quad\mbox{see }\eqref{solution3}\\
\rho^{\star}_{IV} (\xi) &\mbox{for } g_3 \leq g_{\mathrm{NS}} \leq g_4=2, \quad\mbox{see }\eqref{solution4}\\
\end{cases}
\end{equation*}
where $g_1 \equiv 2 - 19/8\sqrt{2}=0.320621 \dots$, $g_2 \equiv (968-499\sqrt{2}+102\sqrt{17})/484 = 1.41088\dots$
and $g_3 \equiv 2 - (9-\sqrt{21})/\sqrt{15(6+\sqrt{21})} = 1.64939\dots $. The emerging physical picture is as follows.
Since $2\int d\xi \xi^2 \rho^{\star}(\xi) = g_{\mathrm{NS}}$, small values
of $g_{\mathrm{NS}}$ are expected to correspond to a large value of
$C_1$ (the strength of the quadratic part of the confining potential $V(\xi)$) and a resulting $\rho^{\star}(\xi)$ that is concentrated near
the left edge $\xi=0$. Making the ansatz
that the density has support on the interval $[0,L_1]$ we determine
it by using Tricomi's formula:
\begin{equation}
\rho^{\star}_{I} (\xi) = \frac{\left(\frac{\sqrt{L_1+1}}{\xi+1} + \frac{C_1}{4}(L_1^2+4L_1\xi-8\xi^2)+a_{I}\right)}{\pi\sqrt{\xi(L_1-\xi)}},
\label{tricomi1}
\end{equation}
where $a_{I}$ is a constant of integration. We now fix $C_1$, $L_1$ and $a_{I}$ by requiring
that $\rho^{\star}_{I}(\xi=L_1)=0$, it is normalized to $1$, and has a second moment
equal to $g_{\mathrm{NS}}/2$. We obtain
\begin{equation}
\rho^{\star}_{I}(\xi) = \frac{\sqrt{L_1-\xi}}{\pi \sqrt{\xi}}\left(\frac{1}{(\xi+1)\sqrt{L_1+1}}+C_1(L_1+2\xi)\right), \label{solution1}
\end{equation}
where $C_1 = \frac{4}{3L_1^2\sqrt{L_1+1}}$ and $1+\frac{5L_1^2-8L_1-16}{16\sqrt{L_1+1}} = g_{\mathrm{NS}}/2$.

For $g_{\mathrm{NS}}>g_1$, $L_1$ becomes greater than $1$, invalidating the solution. This corresponds to a phase transition
in the Coulomb gas: the external potential becomes weak
enough that the density is spread out over the entire available
space to minimize the effects of the inter-particle repulsion. In this
extended phase, $\rho^{\star}$ has support over the entire interval $\xi\in [0,1]$
and is obtained by simply setting $L_1=1$ in \eqref{tricomi1}.
Fixing the integration constant and $C_1$, we obtain
\begin{equation}
\rho^{\star}_{II} (\xi) = \frac{1}{\pi\sqrt{\xi(1-\xi)}}\left(\frac{\sqrt{2}}{\xi+1} + \frac{C_1}{4}(1+4\xi-8\xi^2)\right), \label{solution2}
\end{equation}
where now $C_1=\frac{32}{9}(2-\sqrt{2} - g_{\mathrm{NS}})$. 
For $g_{\mathrm{NS}} > g_2$, $\rho^{\star}_{II}$ goes negative in the middle of
its support, thereby invalidating this solution. 

For $g_2<g_{\mathrm{NS}}<g_3$, we find that no single support solution
is able to satisfy all the constraints on the equilibrium density. In this narrow region, the external potential pushes the Coulomb fluid to the right
edge $\xi=1$ ($C_1$ is {\em negative} for these values of $g_{\mathrm{NS}}$) but {\em cannot fully overcome} the effects of the interparticle
Coulomb repulsion. As a result the Coulomb gas breaks up in this
novel intermediate phase into {\em two} spatially disjoint fluids
separated
by an empty region in the middle.
More precisely, we find using a more general Riemann-Hilbert ansatz \cite{details} that
the solution in the regime $g_2 < g_{\mathrm{NS}} < g_3$ has
two supports, the first on the interval $[0,L_2]$, and the
second on the interval $[L_3,1]$, with $L_3 > L_2$, with the equilibrium
density in these two intervals being given by the formula
\begin{equation}
\rho^{\star}_{III}(\xi) = \frac{-2C_1 \sqrt{(\xi-L_2)(\xi-L_3)^3}(\xi + \frac{(L_2+3L_3+1)}{2})}{\pi\sqrt{\xi(1-\xi)}(1+\xi)},
\label{solution3}
\end{equation}
with $L_3$ related to $L_2$ via the constraint
$ 5-2L_2-6L_3-3L_2^2-6L_2L_3-15L_3^2 = 0
$,
and $L_2$ and $C_1$ being fixed by normalization and second moment equal to $g_{\mathrm{NS}}/2$ (see \cite{supp} for details).

Finally, as $g_{\mathrm{NS}} \to g_3$, $L_2 \to 0$ and $C_1$ is now large enough in magnitude and negative
in sign, giving
way to a conventional single-support solution on
$[L_4,1]$ when $g_{\mathrm{NS}}> g_3$. In this case, Tricomi's formula along with normalization condition yields
\begin{align}
\nonumber\rho^{\star}_{IV}(\xi) &= \frac{\sqrt{2}}{\pi} \frac{\sqrt{\xi - L_4}}{\sqrt{1+L_4}}\frac{1}{\sqrt{1-\xi}} \times\\
&\times\left(\frac{4(2\xi+L_4-1)}{(1-L_4)(1+3L_4)} - \frac{1}{1+\xi}\right), 
\label{solution4}
\end{align}
where $L_4$ is determined by
\begin{equation}
\frac{\sqrt{2}(1-L_4)(1-18L_4-15L_4^2) }{16\sqrt{1+L_4}(1+3L_4)}= \frac{g_{\mathrm{NS}}}{2} - 1. 
\end{equation}

Inserting the analytical expressions 
of the densities in the four phases into the action \eqref{action}, the rate function $\mathcal{R}(g_{\mathrm{NS}})$
can now be evaluated in terms of elementary integrals \cite{details}. This is
shown in Fig.~\ref{Fig1}, where
we display $\mathcal{R}(g_{\mathrm{NS}})$, along with an inset showing the 
analytically calculated curves and Monte-Carlo data for the typical
form of the equilibrium density in each of the four phases. 
Finally, a straightforward asymptotic expansion of these results allows
us to obtain closed
form expressions for the power-law asymptotics of $\mathcal{R}(g_{\mathrm{NS}})$ 
as detailed in the introduction.

\textit{Summary} - In summary, the Coulomb gas formulation of the problem of 
Andreev conductance distribution reveals a rich thermodynamic behavior which can 
be addressed analytically. Four zero-temperature phases in the associated Coulomb 
fluid, dictated by the precise value of $g_{\mathrm{NS}}\in [0,2]$ correspond to 
as many regions in the rate function domain within which 
$\mathcal{R}(g_{\mathrm{NS}})$ is smooth. The central Gaussian region is flanked 
by long-power-law tails with a novel intermediate regime corresponding to a 
disconnected support in the Coulomb fluid density. 
Our result for the full probability distribution
of the Andreev conductance, besides solving a challenging problem, has clear 
physical and 
experimental significance. Such rate functions in related Coloumb gas systems have 
been recently measured experimentally~\cite{Fridman}. 
A direct 
experimental 
confirmation of our predictions in the Andreev case may be within reach with 
existing device setups. Extensions to the case of broken TR appear very 
challenging and are left as 
an open question.


We acknowledge computational resources of TIFR, 
as well as funding
from the Indian DST grants DST-SR/S2/RJN-25/2006 (KD) and DST-SR/S2/RJN-23/2006 (VT), and the Madan Lal Mehta Memorial Trust (SNM).

\end{document}